\documentclass[aps,prl,floats, twocolumn,noshowpacs,superscriptaddress]{revtex4}

\usepackage{graphicx,epsfig}
\usepackage{times}
\usepackage{amssymb,amsmath,multirow,rotate,color}

\begin{document}

\title{Explosive transitions induced by interdependent
  contagion-consensus dynamics in multiplex networks}

\author{D. Soriano-Pa\~nos} 
\affiliation{GOTHAM lab, Institute for Biocomputation and Physics of Complex Systems (BIFI), University of Zaragoza, 50018 Zaragoza, Spain}  
\affiliation{Departamento de F{\'i}sica de la Materia Condensada, Universidad de Zaragoza, 50009 Zaragoza, Spain}

\author{Q. Guo} 
\email{quantongg@buaa.edu.cn}
\affiliation{School of Mathematics and Systems Science, Beihang University \& Key Laboratory of Mathematics 
Informatics Behavioral Semantics (LMIB), Beijing 100191, China}
\affiliation{China Construction Bank, Beijing 100033, China}

\author{V. Latora}
\email{v.latora@qmul.ac.uk}
\affiliation{School of Mathematical Sciences, Queen Mary University of London, London E1 4NS, United Kingdom}
\affiliation{Dipartimento di Fisica ed Astronomia, Universit\'a di Catania and INFN, Catania I-95123, Italy} 
\affiliation{Complexity Science Hub Vienna (CSHV), Vienna, Austria}

\author{J. G{\'o}mez-Garde\~{n}es}
\email{gardenes@unizar.es}
\affiliation{GOTHAM lab, Institute for Biocomputation and Physics of Complex Systems (BIFI), University of Zaragoza, 50018 Zaragoza, Spain}  
\affiliation{Departamento de F{\'i}sica de la Materia Condensada, Universidad de Zaragoza, 50009 Zaragoza, Spain}

\date{\today}

\begin{abstract}
We introduce a model to study the delicate relation between the
spreading of information and the formation of opinions in social
systems. 
For this purpose, we propose a two-layer multiplex network model
in which consensus dynamics takes place in one layer while 
information spreading runs across the other one. The two dynamical
processes are mutually coupled by considering that the control parameters that
govern the dynamical evolution of the state of the nodes inside each
layer depend on the dynamical states at the other layer. In particular,
we explore the scenario in which consensus is favored by the common
adoption of information while information spreading is boosted between
agents sharing similar opinions. Numerical simulations together
with some analytical results point out that, when the
coupling between the dynamics of the two layers is strong enough, a
double explosive transition, {\it i.e.} an explosive transition both for consensus dynamics and
for the information spreading appears. Such explosive transitions lead to bi-stability
regions in which the consensus-informed stated and the
disagreement-ignorant states are both stable solutions.

\end{abstract}


\maketitle
\section{INTRODUCTION}
The  functioning of a wide range of physical, biological, and social complex systems, is often governed by the onset collective properties such as synchronization~\cite{syncrep}, epidemics~\cite{epirep} or the emergence of norms and cooperation ~\cite{castellano} among others. In the last two decades, a number of works have analyzed the role of the networked structure of interactions among the constituents in the emerging collective dynamics of complex systems~\cite{watts,barev,newman,report}. Our understanding of the fundamental mechanisms driving these phenomena is of utmost importance as it provides a solid basis for modeling, predicting, and controlling real dynamical systems~\cite{PastorSatorras2001PRL,Nishikawa2003PRL,JGGPRL}. 

In the recent years, network and complexity sciences have moved one
step forward in this direction by considering that, very frequently,
the elements of many real complex systems are subject to different dynamical
interactions at the same time. Moreover, in most of the cases, these
interactions depend on each other thus connecting the different
dynamical processes that occur simultaneously. Examples of the
coexistence and non-trivial interdependence between two or more 
dynamical processes are very common in social systems and in the natural
sciences. For instance,
human prevention behavior and epidemic spreading
\cite{vacc1,vacc2,vacc3,vacc4} or the structural-functional relationships
within cortical areas in the brain \cite{Bullmore}. The study of such
coupled dynamical processes has been largely spurred by the
introduction of novel frameworks to deal with 
multiplex networks~\cite{Regino,DeDomenico2013PRX,Kivela2014JCN,Boccrev}. In fact,
multiplex networks are the natural way to model the existence of different
dynamical interactions among the same set of units~\cite{Reinares,Diffusion,Cozzo,Granell,Gambuzza,DelGenio,Amato,Antonioni,Nicosia,moreno2017,Metamultiplex}.

In this work we introduce and study a model of two coevolving
socially-inspired processes: formation of opinions and information
spreading. These two dynamics are mutually coupled in such a way that
the transmission of information from a spreader agent to a receiver is
boosted when the neighbors of the latter share similar opinions ~\cite{Lotero}. In
addition to this, the alignment of the opinion of an agent to those
of her neighbors is fostered when such neighbors spread the voice
simultaneously. The model, and in particular the adopted type of 
interdependence of the two processes, captures everyday
life examples in which the use of technology or the adoption of new
ideas by an individual happens in virtue of the consensus found among
her acquaintances \cite{CC1} and, in turn, the common adoption of
these novelties boosts the degree of homophily needed for the creation
of social consensus \cite{Axelrod}. 

Our results point out that the interplay between opinion and spreading dynamics may dramatically
alter the critical properties of their associated transitions leading to abrupt onsets of epidemic and consensus. It is remarkable, that these explosive transitions are not produced by using any of the ingredients usually found in single networks~\cite{physrepexp} but they are the result of the coupling induced by the multiplex architecture. The existence of these explosive transitions lead to the appearance of bistable regions where the multiplex network
can switch between active to inactive dynamical phases due to small perturbations.

The rest of this work is organized as follows: In Sec.~\ref{sec:II} we characterize the dynamical coupling between layers. Then, in Sec.~\ref{sec:III} we illustrate numerically the onset of explosive consensus and contagion transitions. These sharp transitions are analyzed in Sec.~\ref{sec:IV} by identifying the critical coupling for which they occur. Finally, in Sec.~\ref{sec:V} we round off by discussing the main results and future directions of our work.



\begin{figure*}[t!]
\centering
\epsfig{file =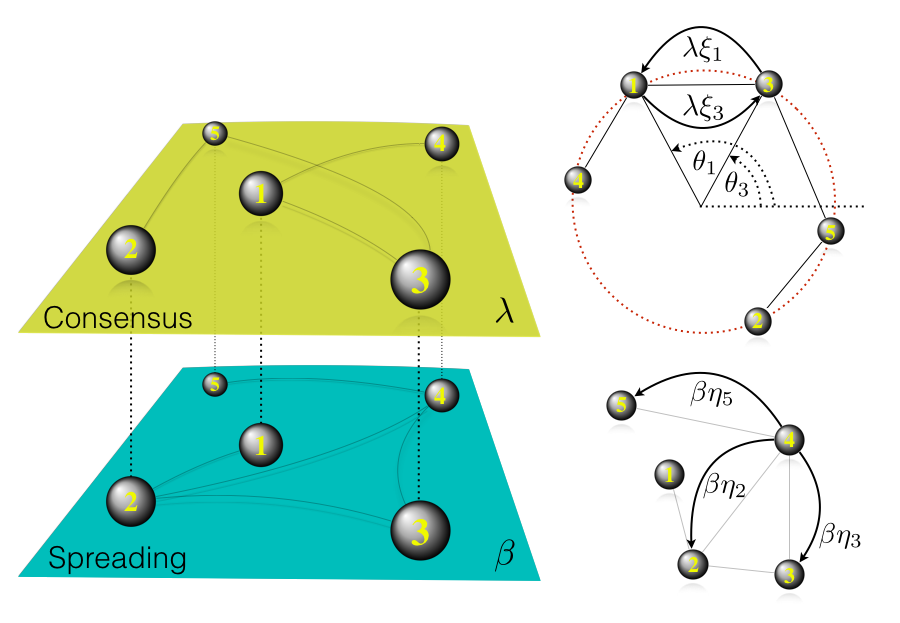, clip =,width=0.62\linewidth }
\caption{Left: Schematic representation of our model on a multiplex
  network with $M=2$ layers and $N =5$ nodes.  The first (top) layer
  accounts for the consensus dynamics, which is modeled by a Kuramoto
  model as in Eq.~(\ref{eq:Kuramoto}), whereas the second (bottom)
  layer describes the spreading of information according to the SIS
  model as in Eq.~(\ref{eq:SIS}). Right: The coupling strength $\lambda$ between opinions (top) as well
as the contagion rate $\beta$ (bottom) have been modified as in
Eq.~(\ref{eq:xi}) and Eq.~(\ref{eq:eta}) respectively in order to mutually
couple the synchronization process to the spreading of information. 
}
\label{fig1}
\end{figure*}

\section{Model of interdependent dynamics} 
\label{sec:II}
 
In order to describe the delicate interplay between information
spreading and the formation of consensus in a social system, we
introduce here a model in which the two processes take place at the
different layers of a multiplex network with $M=2$ layers, and are
mutually coupled. We deal with a multiplex network following the
assumption that there exists a one-to-one correspondence between nodes
(the agents) in different layers, so that each layer is composed by
the same set of $N$ nodes.  However, the topologies of the two layers
can in general be different and are described by the adjacency
matrices $A^{[1]} = \{ a^{[1]}_{ij} \} $ and $A^{[2]} =\{ a^{[2]}_{ij}
\}$ respectively. These matrices are defined such that $a^{[1]}_{ij}=1$ ($a^{[2]}_{ij}=1$) if a
link exists between nodes $i$ and $j$ in the first (second) layer, and
otherwise $a^{[1]}_{ij}=0$ ($a^{[2]}_{ij}=0$).  We denote the degree
of node $i$ in the first (second) layer as $k_i^{[1]}=\sum_{j=1}^N
a^{[1]}_{ij}$ ($k^{[2]}_i=\sum_{j=1}^N a^{[2]}_{ij}$).

Our model can be phrased in terms of a general formalism for interdependent dynamical networks proposed in Ref.~\cite{Nicosia}. If we denote respectively as ${\bf x(t)} = \{x_1(t), x_2(t),\ldots, x_N\} \in \Re^N$ 
and ${\bf y(t)} = \{y_1(t), y_2(t),\ldots, y_N\}\in \Re^N$ the states of the nodes at the two layers,  
the evolution of the system can be written as:  
\begin{equation}
    \left\{ \begin{array}{c}
       \dot{x}_i = F_{\xi_i} ( {\bf x} , A^{[1]} )   \\
       \dot{y}_i = G_{\eta_i}  ( {\bf y} , A^{[2]} )     \\
\end{array}
\right.
\qquad 
i=1,2,\ldots N
\label{eq:gen}
\end{equation}
where the dynamics of state $x_i$ ($y_i$) of node $i$ in the first (second) layer is governed by a function $F_{\xi}$ ($G_{\eta}$) of the dynamical state ${\bf x}$ (${\bf y}$) and of the structure $A^{[1]}$ ($A^{[2]}$) of the first (second) layer.
Notice that, following Ref.~\cite{Nicosia}, functions $F_{\xi}$ and $G_{\eta}$ in our model are taken to be dependent on the parameters $\xi$ and $\eta$, and this is the key ingredient to connect the two dynamical processes. Namely, we assume that the parameter $\xi_i$ of function $F_{\xi_i}$ at the first layer is itself a function of time depending on the dynamical states $\{y_j(t)\}$ at the second layer of the neighbors of node $i$ at the first layer ($a^{[1]}_{ij}=1$). Analogously, the evolution of the parameter $\eta_i$ at the second layer depends on the states $\{x_j(t)\}$ at the first layer of the neighbors of node $i$ at the second layer ($a^{[2]}_{ij}=1$). In this way, the system of Eq.~(\ref{eq:gen}) is completed by the following system of equations:
\begin{equation}
    \left\{ \begin{array}{c}
   \xi_i (t)= f (\{y_j(t) | a^{[1]}_{ij}=1\} ) \;
 \\
   \eta_i (t) = g (\{x_j(t) | a^{[2]}_{ij}=1\}) \;
 \end{array}
\right.
\qquad 
i=1,2,\ldots N
\label{eq:gen2}
\end{equation}
once the two functions $f$ and $g$ are assigned. 

As illustrated in Fig.~1, the first layer in our model accounts for the dynamics underlying the formation of consensus in a social system, while the second layer describes the contagion processes mimicking
the spread of ideas/products. The dynamical state $x_i(t)$ of node $i$ at the first layer represents the
opinion of individual $i$. This opinion is described by means of a phase variable, i.e.  
$x_i(t)=\theta_i(t)\in[-\pi,\pi]$. The time evolution of $x_i(t)$ is modeled via the Kuramoto model of coupled
phase-oscillators~\cite{Kuramoto75,STR,Conrad}, so that the first
set of equations in the system of Eq.~(\ref{eq:gen}) reads:
\begin{equation}
  \dot{\theta}_i(t)=F_{\xi_i}({\bf \theta}, A^{[1]})=\omega_i +
  \lambda \xi_i(t)  \sum_{j=1}^N a^{[1]}_{ij}\sin \left[ \theta_j(t) -\theta_i(t) \right]
  \label{eq:Kuramoto}
\end{equation}
where $\omega_i$ is the natural frequency of node $i$. Notice that $\lambda$ is a global coupling strength, while the local coupling strength associated to node $i$ is  modulated by the dynamical variable $\xi_{i}(t)$
that changes in time depending on the  dynamics of the second layer, as sketched in Eq.~(\ref{eq:gen2}), in a way that will be specified below.

The dynamical state $y_i(t)$ of node $i$ at the second layer
represents the probability of node $i$ of being active as
user/spreader of an idea, namely $y_i(t)= p_i(t) \in [0,1]$.
The time evolution of $p_i(t)$ is modeled through a 
Susceptible-Infected-Susceptible (SIS) model. In this way we identify
susceptible (ignorant) agents as those who do not transmit any information,
whereas the infected ones correspond to active users (spreaders)
who disseminate the information to the rest of the population. 
Under this framework, a susceptible (S) that has a spreader
neighbor can be infected by it at time $t$ through the process
$S+I\rightarrow 2I$ and become spreader (I) with a probability
$\beta\eta_i(t)$. In addition, a spreader can return to its ignorant
state through the process $I \rightarrow S$ with a probability
$\mu$. Such a dynamics can be cast in the form of a Markovian
evolution for the probability $p_i(t)$ that a node $i$ is spreader at
time $t$ as \cite{gomezEPL,Guerra,gomezPRE}:
\begin{equation}
 \dot{p}_i(t)=-\mu
 p_i(t)+(1-p_i(t))\left[1-\prod_{j=1}^{N}\left(1-a^{[2]}_{ij}\beta\eta_i(t)p_j(t)\right)\right]\;
 \label{eq:SIS}
\end{equation}
Notice that, at variance with the usual SIS model, here the
microscopic contagion probability $\beta\eta_i(t)$ of node $i$ may
differ from node to node and, also, it changes in time due to the
presence of factor $\eta_i(t)$, in close analogy with the presence of
factor $\xi_i(t)$ in the effective coupling of unit $i$ in the
consensus layer.
 
Finally, we need to assign the time-dependent functions $\{\xi_i(t)\}$
and $\{\eta_i(t)\}$ that mutually couple the consensus dynamics and
the process of contagion as sketched in Eq.~(\ref{eq:gen2}). 
In order to define $\eta_i(t)$, we need to
capture the influence that consensus at layer 1 has on the contagion
dynamics at layer 2. With this purpose we evaluate the local degree of
consensus $r_i(t)$ around node $i$ at time $t$ by considering the
values of $\theta_j(t)$ in the neighborhood of node $i$.
Notice, however, that the neighbors of node $i$ are taken 
in the second layer (where information spreading takes place), i.e. by
using the adjacency matrix $ \{ a^{[2]}_{ij} \} $, since it is the consensus among potential spreaders what facilitates the transmission of ideas. The local degree of consensus of node $i$ is defined as the modulus of
the complex function:
\begin{equation}
r_i(t)e^{i\psi_i(t)}=\frac{1}{k^{[2]}_i}\sum_{j=1}^Na^{[2]}_{ij}e^{i\theta_j(t)}\;
\label{eq_ri}
\end{equation}
so that we get $r_i\simeq 0$ in the absence of local consensus and $r_i=1$
otherwise. Once evaluated $r_i(t)$, we can write the second of 
Eq.~(\ref{eq:gen2}) as: 
\begin{equation}
\eta_i(t)=\frac{1}{1+\exp[-\alpha(r_i(t)-r^{*})]}\;
\label{eq:eta}
\end{equation}
The use of the Fermi function with a tuning parameter $\alpha>0$ implies that, for large enough values of $\alpha$, when  $r_i(t)\rightarrow 0$, i.e. when the local consensus around $i$ is small, the contagion probability towards $i$, $\beta\eta_i(t)$, tends to $0$. On the other hand, when consensus among the neighbors of $i$ increases, their influence over $i$ also grows, approaching $\beta$ as $r_i(t)\rightarrow 1$. This way $r^{*}$ acts as a threshold so that for $r_i(t)>r^{*}$ ($r_i(t)<r^{*}$) we have $\eta_{i}(t)>0.5$ ($\eta_{i}(t)<0.5$). For the sake of simplicity, in the following we set $r^{*}=0.5$.

Lastly, we model the influence that the contagion dynamics of layer 2 has on the formation of consensus in layer 1. To this aim, the node-depending coupling constant $\lambda\xi_i(t)$ of the Kuramoto model at layer $1$ is chosen to be dependent on the number of spreaders around $i$ at layer $1$. Specifically, $\xi_i(t)$ is defined as the fraction of spreaders among the neighbors of node $i$ in layer $1$, so that the first of
Eq.~(\ref{eq:gen2}) reads: 
  \begin{equation}
\xi_i (t) = \frac{ \sum_{j=1}^N  a_{ij}^{[1]}  p_j(t) }{k^{[1]}_i} \;
\label{eq:xi}
\end{equation}

Summing up, in our model the state $(\theta_i(t), p_i(t))$ of each node $i$, with $i=1,2,\ldots,N$, evolves in time as in Eq.~(\ref{eq:Kuramoto}) and  Eq.~(\ref{eq:SIS}), where the two parameters $\xi_i$ and $\eta_i$  depend in turn on the state $(\theta_i(t), p_i(t))$ as in Eq.~(\ref{eq:xi}) and Eq.~(\ref{eq:eta}), mutually coupling the two dynamical  processes. Notice that in this way both the infection probability $\beta\eta_i$ and the Kuramoto coupling strength $\lambda\xi_i$ of a node $i$ are obtained by taking average over the neighbors in the layer
that governs the corresponding dynamics, {\em i.e.}, layer $1$ for $\xi_i$ and layer $2$ for $\eta_i$. However, the averaged dynamical quantities correspond to the node states at the other layer, {\em i.e.} the phases for $\eta_i$ and the probabilities of being infected for $\xi_i$, thus closing the feedback loop between spreading and consensus dynamics. The way these two processes interplay follow, as discussed above, the rationale that the existence of consensus facilitates the adoption of ideas and that it is the simultaneous spread of ideas what foster the alignment of opinions.

\section{RESULTS}
 \label{sec:III}
In order to characterize the effects of the interplay between spreading and consensus dynamics, we explore the dynamical behavior of the coevolving model focusing on the synchronization and epidemic onsets. To this aim, we start by infecting a little fraction $\rho$ of agents and by setting randomly the oscillator phases $\theta_i$ within a range $\theta_i \in (-\pi,\pi]$. The natural frequencies of oscillators $\{w_i\}$ are also randomly chosen within $w_i \in [-0.5,0.5]$. We take $\lambda$ and $\beta$ as the natural control parameters for Kuramoto and SIS dynamics respectively. The order parameters are also the usual ones for both dynamical systems. Namely, the degree of global consensus is measured by using the Kuramoto order parameter $r$ defined by the 
complex number: 
\begin{equation}
r(t)e^{i\psi(t)}=\frac{1}{N}\sum_{j=1}^Ne^{i\theta_j(t)}\;,
\end{equation}  
which represents the centroid of all oscillators placed on the complex unit circle. In its turn, for the SIS model we monitor the evolution of the fraction of infected individuals:
\begin{equation}
I(t)=\frac{1}{N}\sum_{j=1}^{N}p_j(t)\;.
\end{equation} 
As usual, the order parameters, $r$ and $I$,  are measured by making a time average of $r(t)$ and $I(t)$, once the stationary regime of the dynamics is reached. To reach this stationary state, we integrate Eq.(\ref{eq:Kuramoto}) by using the fourth order Runge-Kutta method and Eq. (\ref{eq:SIS}) using an Euler method, both with time steps $\delta t=0.01$.

The networks used to build the multiplex configurations are random Erd\"os-Renyi (ER) and scale-free (SF) networks with $N=500$ nodes and average degree $\langle k\rangle \simeq 4$. The use of these two topologies allows us to study the role of degree heterogeneity in the evolution of consensus and spreading dynamics.
As anticipated above, we denote the multiplex considering that the first layer contains consensus dynamics whereas information spreading takes place on top of the second one.

\begin{figure}[t!]
\centering
\epsfig{file =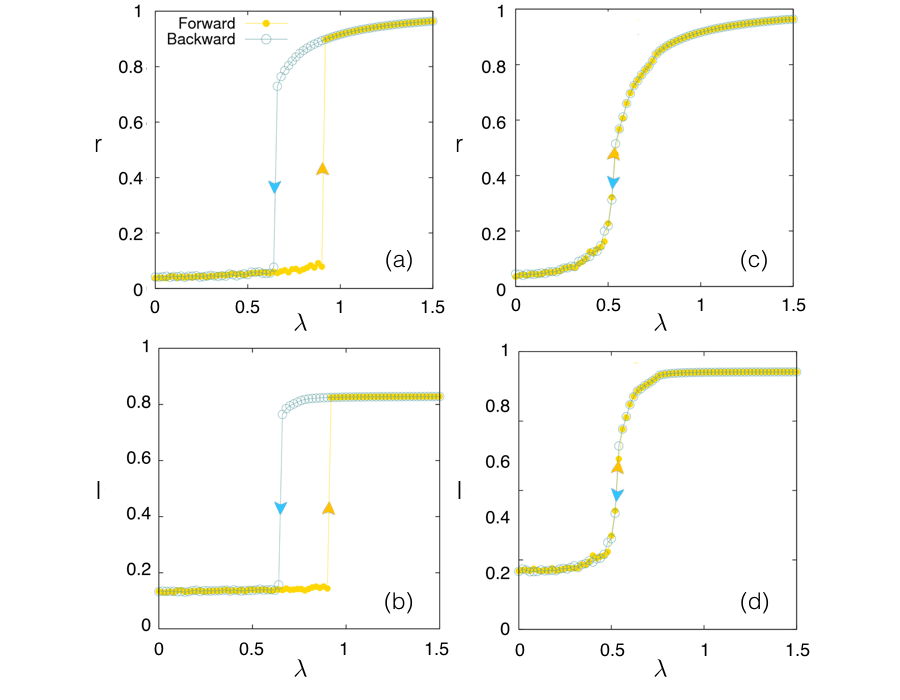, clip =,width=1.05\linewidth }
\caption{Average global consensus $r$ (Top) and fraction of information spreaders $I$ (Bottom) as a function of the coupling constant $\lambda$ for the SF-ER multiplex configuration. These order parameters have been computed adiabatically by increasing the value of the coupling constant $\lambda$ (Forward)  from $\lambda=0$ or by decreasing it (Backward) from $\lambda=1.5$. The contagion rate values used are for (a) and (b) $\beta=0.70$ and for (c) and (d) $\beta=1$. The rest of the model parameters are set to $\alpha=10$ and $\mu=1.0$.} 
\label{fig2}
\end{figure}

In Fig.~\ref{fig2}, we have computed the diagrams for global consensus and fraction of infected people using
a SF-ER multiplex by keeping fixed the contagion probability, $\beta$, in the ER layer and varying the consensus coupling in the SF one. To this aim, we have computed the forward (increasing $\lambda$) and backward (decreasing $\lambda$) diagrams. Panels (a)-(b) and (c)-(d) show drastically different transitions. On one hand, (a)-(b), that correspond to $\beta=0.70$, show an abrupt transition both for the degree of consensus and the fraction of spreaders. These diagrams are characterised by the existence of regions of bistability where the solutions corresponding to absence of global consensus and information spreaders coexist with those displaying macroscopic coherence and spreading. On the contrary, in panels (c)-(d), corresponding to $\beta=1$, show a smooth and continuous transition, {\it i.e.} the expected onset from the usual Kuramoto and SIS models. As we show below the particular type, smooth of explosive, of transition depends on the multiplex configuration and on the $\beta$ value.

\begin{figure}[t!]
\centering
\includegraphics[width=1.0\columnwidth]{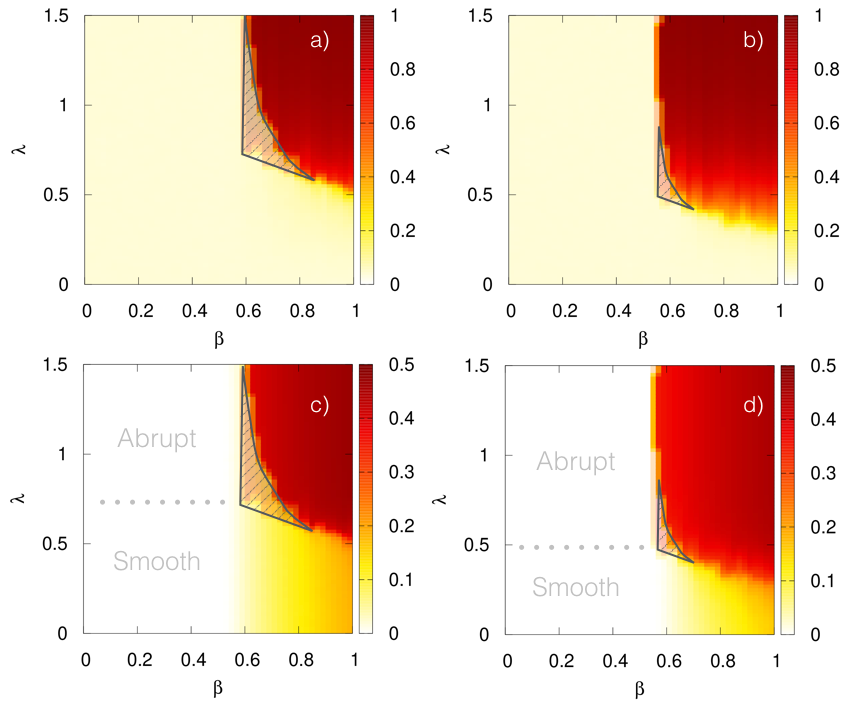}
\caption{Diagram of the average global consensus $r$ (top) and fraction of spreaders nodes $I$ (bottom) as a function of the infectivity $\beta$ and the coupling parameter $\lambda$ for SF-ER (left) and SF-SF (right) multiplex networks. Both magnitudes have been obtained by averaging their values during $T=400$ steps. The shadowed regions in the panels contain the areas of the space of parameters ($\lambda,\beta$) where hysteresis cycles appear due to the coexistence of two stable solutions: total consensus-disagreement for the synchronization dynamics and active-inactive spreaders from the point of view of the information spreading process.} \label{fig3}
\end{figure}

To have a broader picture about the phenomenon described above, in Fig.~\ref{fig3}, we represent the diagrams for global consensus and fraction of infected people as a function of both $\beta$ and $\lambda$ for SF-ER (left panels) and SF-SF (right panels) multiplexes. At first sight, for both topologies, we observe that below a critical value $\beta_c$, the single stable solution is the absence of global consensus and information spreaders. Above this threshold, we can find different stable solutions depending on the Kuramoto coupling constant $\lambda$. 
Namely, for small values of $\lambda$ below a certain threshold $\lambda_c$, the stable solution is the absence of consensus and the presence of a small fraction of infected people $I$ which depends on the value of the contagion rate $\beta$. 
Above this threshold $\lambda_c$, the transitions observed when increasing $\beta$ are totally different turning abrupt at a value $\beta_c$. Interestingly, the value $\beta_c$ where the abrupt onsets appear is the same as that $\beta_c$ found when the transition is smooth ({\em i.e.} when $\lambda<\lambda_c$). Interestingly, in the regime $\lambda>\lambda_c$ the abrupt transitions incorporate bistability regions (see stripped areas in Fig.~\ref{fig3}) where the coexistence of two solutions (corresponding to large and small order parameters) explains the hysteresis cycles shown in Fig.~\ref{fig2} when $\lambda$ is varied using a fixed value of $\beta\gtrsim\beta_c$.

Once described the diagrams in the $(\beta,\alpha)$-plane let us identify the main differences between ER-SF and SF-SF multiplexes. By comparing panels (a)-(c) and (b)-(d) in Fig.~\ref{fig3}, it is clear that the value of the critical coupling $\lambda_c$ (separating the regions corresponding to smooth and explosive transitions) is lower for the SF-SF configuration than for the SF-ER one. To explain this, we must take into account that both SF layers in the SF-SF configuration are positively correlated so that hubs promote the interplay between consensus dynamics and information spreading, thus anticipating the explosive onsets. Another remarkable difference between both multiplexes, is that the bistable regime is hindered in the SF-SF configuration with respect to the SF-ER one. 


At this point, we can understand the role that each process plays on the intertwined dynamics. It becomes clear that the epidemics behaves as the limiting process, for the emergence of consensus requires the existence of active spreaders but no vice-versa. In its turn, the synchronization dynamics, monitored by the coupling constant $\lambda$, behaves as an external force which drives the system from a practically inactive phase to an active one. 

\section{ESTIMATION OF THE GLOBAL THRESHOLD}
\label{sec:IV}
To support the former results, we now derive the contagion threshold $\beta_c$ of the multiplex. As the epidemic state appears smoothly despite the absence of consensus, for the sake of simplicity we make this derivation assuming that $\lambda =0$. This allow us to consider $r_i (t)$ as a constant, $r_i(t)\simeq r_i$,  and to decouple Eqs. (\ref{eq:Kuramoto},\ref{eq:SIS}). Then we suppose that the spreading dynamics has reached the stationary state so that $\dot{p_i} =0 \ \forall i$. Therefore, Eqs.(\ref{eq:SIS}) read as:
\begin{eqnarray}
\mu p_i^* = (1-p_i^*)\left[1-\prod_{j=1}^{N}\left(1-a^{[2]}_{ij}\frac{\beta}{1+e^{-\alpha(r_i-0.5)}}p_j^*\right)\right]\ .
\label{eq:stationary}
\end{eqnarray}
In the stationary regime and close enough to the epidemic threshold the individual probabilities of being spreader are very small but no zero, {\it i.e.}, $p_i^*=\epsilon_i \ll 1$. This way, one can linearize Eq.~(\ref{eq:stationary}) that, neglecting term orders higher than $\epsilon$, turns into:
\begin{equation}
\mu \epsilon_i = \sum\limits_{j=1}^N\underbrace{\frac{1}{{1+e^{-\alpha(r_i-0.5)}}}a^{[2]}_{ij}}_{{\cal M}_{ij}}\epsilon_j\ .
\end{equation}
The former expression is an eigenvalue problem, being the epidemic threshold the minimum value of $\beta$ compatible with this equation. Therefore, the threshold is related to the maximum eigenvalue of the epidemic layer as:
\begin{equation}
\beta_c = \frac{\mu}{\Lambda_{max}\left(\cal M\right)}\ .
\end{equation}

The epidemic threshold is the same as in the usual SIS model \cite{gomezEPL} but shifted by the influence of the local consensus. In this sense, it can be initially thought that the absence of global consensus (recall that agents opinions are placed following a random distribution) and opinion alignment ($\lambda$=0) leads to a zero value of the local consensus. This would imply that, for large $\alpha$ values, the epidemic state would not be a stable solution for physically meaningful $\beta$ values and, consequently, it would not be observed. 

To explain the observation of an epidemic threshold, we must take into account that we are dealing with {\it sparse} networks, in which the number of neighbors of each node is significantly smaller than the network size. As a consequence, the computation of the local consensus according to Eq. (\ref{eq_ri}) only involves a few nodes, what leads to its non-zero value for each oscillator despite the initial random conditions. To illustrate this argument, we have considered the case in which the information spreading layer is the ER network. There, we can approach, due to the almost homogeneous degree distribution, the local consensus for each node to its average. This way, the epidemic threshold can be expressed as:
\begin{equation}
\beta_c (\text{SF-ER}) \simeq \frac{1+e^{-10(\langle r\rangle - 0.5)}}{\Lambda_{max}(\bf a^{[2]})} = 0.52\;,
\label{ecuacion}
\end{equation}
that, as observed from Fig.~\ref{fig3}, constitutes a good estimation. 

It is also remarkable that for the SF-SF multiplex the threshold $\beta_c$ is roughly the same than that corresponding to the SF-ER case. This is a surprising result since, in usual epidemiological models, the high degree heterogeneity of SF networks leads to an epidemic threshold much smaller than that of ER ones with the same average degree. However, in the proposed intertwined dynamics, the existence of a large number of neighbors implies the access to a wide set of opinions, thus reducing the value of the local consensus. Thus, although $\Lambda_{max}({\bf a}^{[2]})$ becomes larger for the SF-SF multiplex, the large value in the denominator of Eq.~\ref{ecuacion} is compensated by the decrease of $\langle r\rangle$ in the numerator as a consequence of the large number of neighbors surrounding hubs described above.  Consequently, this balance causes that the threshold $\beta_c$ takes similar values for SF-ER and SF-SF multiplexes.

\section{CONCLUSIONS}
 \label{sec:V}
In this work we have studied a model which allow us to quantify the effects of the interplay between information spreading and the emergence of social consensus. To do so, we have relied on a two layer multiplex framework in which one layer codifies the information spreading according to a SIS model and the other one contains the opinion dynamics dictated by a Kuramoto model. To couple both dynamics, we have considered that the control parameters of each dynamics depend on the dynamical state of the other process. Specifically, we assume that both processes enhance mutually in such a way that the presence of a lot of spreaders foster global consensus and the other way round, {\it i.e.}, the existence of similar opinions about ideas promotes their adoption.

Results of numerical simulations have revealed usual features about information spreading and consensus inside populations. For instance, we have characterized the limiting role of information spreading, since the absence of spreaders of an idea impedes the achievement of a global consensus on it among people with different initial opinions. On the other hand, we have also observed that the direct correlation between information spreading and opinion alignment leads to the onset of abrupt transitions. These explosive transitions are very relevant due to the drastic changes induced by perturbations in the bi-stability regions. This finding adds to recent studies devoted to determine the conditions which lead to explosive phenomena in monolayer networks \cite{GomezGardenes2011PRL,Science,Motter2013NaturePhys,Buldyrev2010Nature,Zhang2014} as well as in multilayer networks \cite{Nicosia,Zhang2015,Danziger}. 

Finally, we have analyzed the influence of the multiplex configuration on the threshold of the proposed intertwined dynamics. For this purpose, we have linearized the model equations unveiling that the degree of local consensus along with the topology of the information spreading layer determine the value of this threshold. In fact, the interplay between these two ingredients leads to a larger value of this threshold for heterogeneous networks than for homogeneous ones. Apparently, this is a counter-intuitive and atypical result, since usually epidemiological models \cite{epirep,epirep2} predict smaller values of the epidemic threshold for heterogeneous networks. However, in heterogeneous networks the interaction with a large number of agents, whose opinions are initially randomly distributed, makes local consensus more difficult to achieve, thus inhibiting the propagation of information. 

In a nutshell, the formalism introduced here constitutes a simple framework to characterize the mutual influence between the propagation of ideas inside a population of agents and the alignment of their opinions. Our results point out that imposing a positive correlation between both processes leads to the emergence of explosive phenomena in both the spreading and consensus dynamics thus providing an alternative and multiplex-based way of creating abrupt onsets in models (here SIS and Kuramoto) with associated smooth transitions.

\acknowledgments  D.S.P. acknowledges financial support from Gobierno de Arag\'on through a doctoral fellowship. Q.G. acknowledges financial support from China Scholarship Council(CSC) (No.201406020055). V.L. acknowledges support from the EPSRC projects GALE, EP/K020633/1, and EP/N013492/1. J.G.G. acknowledges support from MINECO through grant number FIS2015-71582-C2-1. J.G.G \& D.S.P. acknowledge support from MINECO and Fondo Europeo de Desarrollo Regional (FEDER) (Grants FIS2015-71582-C2 and FIS2017-87519-P), and Gobierno de Arag\'on/Fondo Social Europeo (Grant E36-17R to FENOL group). 

\end{document}